\documentclass[a4paper,
               ]{jacow}
\usepackage{pdfpages,multirow,ragged2e} %
\makeatletter%
	\ifboolexpr{bool{xetex}}
	 {\renewcommand{\Gin@extensions}{.pdf,%
	                    .png,.jpg,.bmp,.pict,.tif,.psd,.mac,.sga,.tga,.gif,%
	                    .eps,.ps,%
	                    }}{}
\makeatother

\ifboolexpr{bool{xetex} or bool{luatex}} 
 {}                                      
 {\usepackage[utf8]{inputenc}}           

\usepackage[USenglish]{babel}

\usepackage{subcaption}

\ifboolexpr{bool{jacowbiblatex}}%
 {%
  \addbibresource{jacow-test.bib}
  \addbibresource{biblatex-examples.bib}
 }{}
\begin{document}

\title{ABEL: The Adaptable Beginning-to-End Linac \\simulation framework}

\author{J. B. B. Chen\thanks{j.b.b.chen@fys.uio.no}, E. Adli, P. Drobniak, O. G. Finnerud, E. H\o{}rlyk, D. Kalvik, C. A. Lindstr\o{}m,\\ F. Pe\~na, K. Sjobak, Department of Physics, University of Oslo, Oslo, Norway}
	
\maketitle

\begin{abstract}
    We introduce ABEL, the Adaptable Beginning-to-End Linac simulation framework developed for agile design studies of plasma-based accelerators and colliders. ABEL’s modular architecture allows users to simulate particle acceleration across various beamline components\*. The framework supports specialised codes such as HiPACE++, Wake-T, ELEGANT, GUINEA-PIG, CLICopti and ImpactX, which facilitate precise modelling of complex machine components. Key features include simplified models for addressing transverse instabilities, radiation reactions, and ion motion, alongside comprehensive diagnostics and optimisation capabilities. Our simulation studies focus on the HALHF plasma linac, examining tolerances for drive beam jitter, including effects of self-correction mechanisms. Simulation results demonstrate ABEL's ability to model emittance growth due to transverse instability and ion motion, highlighting the framework’s adaptability in balancing simulation fidelity with computational efficiency. The findings point towards ABEL’s potential for advancing compact accelerator designs and contribute to the broader goals of enhancing control and precision in plasma-based acceleration.
\end{abstract}

\section{INTRODUCTION}
Plasma-based accelerators (PBAs) offer a promising path toward compact, high-gradient particle acceleration, potentially enabling next-generation linear colliders and light sources \cite{Veksler, Tajima_LWFA_1979, Ruth_1984, Chen_PWFA_1985}. However, the complexity of plasma dynamics, coupled with the diversity of components in a full accelerator chain, makes accurate modelling of these machines both challenging and computationally intensive. While experimental efforts continue to provide valuable insights, they remain costly and limited in scope, reinforcing the need for self-consistent simulation tools that can guide design and optimisation strategies for both PBA machines and experiments.

A plasma-based linac or collider consists of many different beamline elements such as beam sources, acceleration stages, interstages and beam delivery systems (BDS). These beamline elements require specialised, often incompatible codes, limiting direct transfer of simulation outputs. In addition, the physics at the interaction point (IP) also need to be well-understood in order to maximise luminosity and minimise unwanted background particles.

To address this gap, we introduce the Adaptable Beginning-to-End Linac simulation framework (ABEL), a modular, object-oriented simulation framework written in Python. ABEL enables self-consistent simulation of plasma wakefield acceleration (PWFA) and collider concepts by linking a suite of specialised codes - such as HiPACE++ \cite{diederichs_hipace_2022}, Wake-T \cite{pousa_waket_2019}, ELEGANT \cite{borland_elegant}, GUINEA-PIG \cite{thesis_Daniel}, CLICopti \cite{Sjobak_clicopti, Ogren_clicopti}, RF-Track \cite{Latina_rftrack} and ImpactX \cite{Huebl_impactx} - using the openPMD-standard \cite{OpenPMD}. Its design allows users to flexibly switch between high-fidelity codes and simplified native models, enabling studies that balance physical accuracy with computational efficiency. ABEL can model full accelerator chains, supports extensive diagnostics, and enables isolating specific physical effects such as transverse instabilities, radiation reaction, and ion motion.

Beyond physics simulation, ABEL also functions as a system-level tool for optimisation and machine design similar to system codes in the nuclear fusion community \cite{Morris_PROCESS_2024, Chan_TESC_2015, Dragojlovic_ARIES}. It can run ensemble simulations with stochastic input variations (e.g., beam jitter), scan multi-dimensional parameter spaces, and perform Bayesian optimisation \cite{Jones_Bayesian_optimisation} of machine parameters. These features make ABEL especially suited for agile design studies of future plasma-based linear accelerators (linacs) and colliders. In this work, we apply ABEL to model the HALHF plasma linac, demonstrating its capabilities in evaluating drive beam jitter tolerances and emittance growth mechanisms.

\section{FRAMEWORK STRUCTURE}
In ABEL, all beamline elements are represented as Python classes, with subclasses providing different levels of fidelity and computational efficiency. Specifically, the beamline elements are structured as parent classes that define fundamental parameters and methods shared by all subclasses. Each subclass incorporates its own methods for simulating a beamline element, spanning from simple idealised models to advanced implementations using specialised codes.

The beamline elements can be assembled into chains to form e.g. linacs and colliders, and ABEL handles all communication and I/O operations between the elements, as well as supporting an extensive suite of diagnostics.

\begin{figure}[b]
    \centering
    \includegraphics[width=0.90\columnwidth]{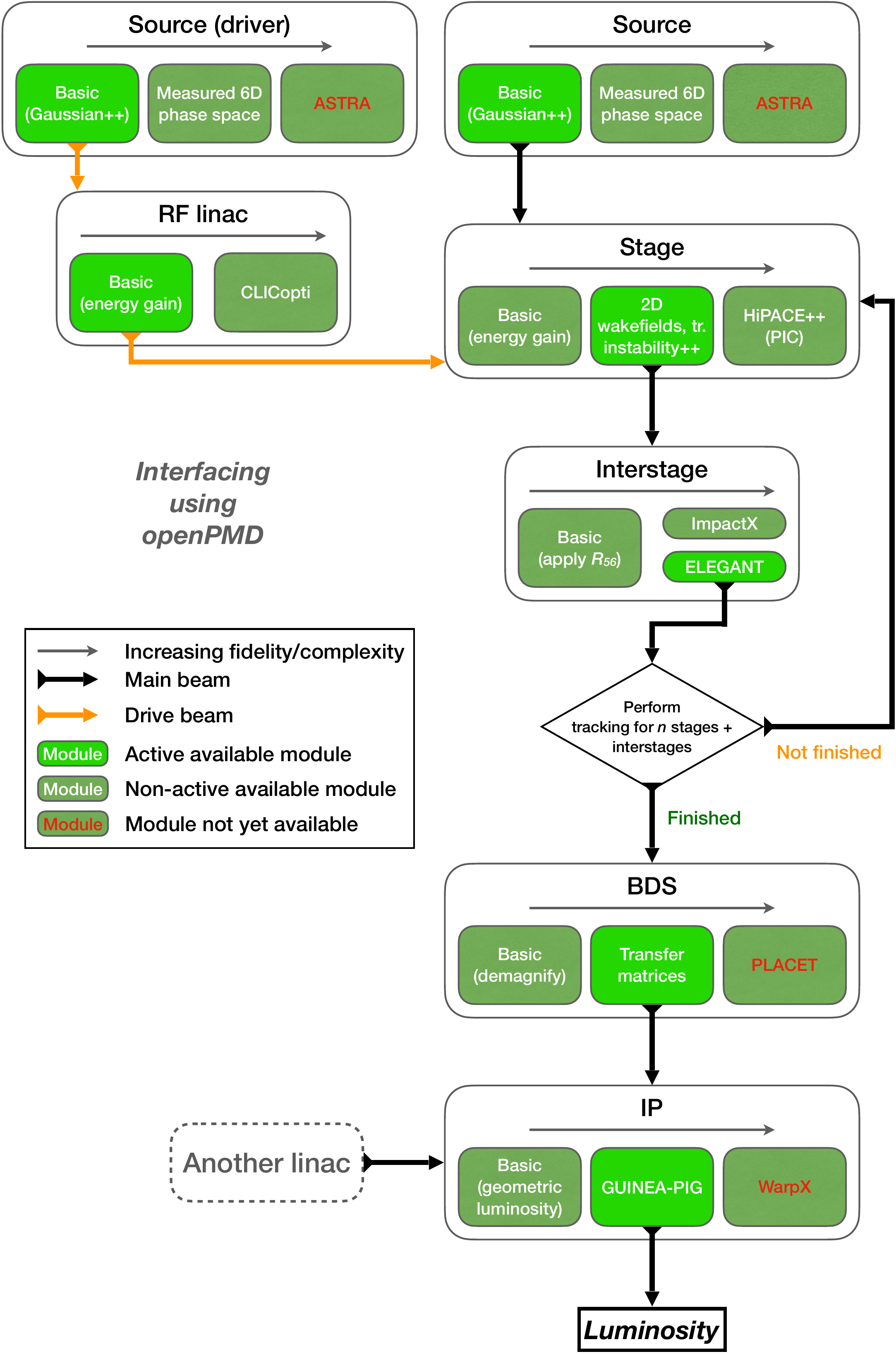}
    \caption{ABEL structure layout and data flow showcased as a PWFA collider. The beamline elements are implemented as Python classes, which all have a selection of subclasses with varying fidelity and speed. The selected modules in this example are marked in bright green.}
    \label{fig:ABEL_structure}
\end{figure}

The framework structure is illustrated in Fig.~\ref{fig:ABEL_structure}, where a generic multi-stage PWFA collider consisting of beam sources, plasma acceleration stages, interstages, BDS and IP is used as an example. Here we will briefly outline the available beamline classes in ABEL, emphasising classes that are relevant for the discussions in the use case section.

\subsection{Beam source}
The particle beams used in the tracking is generated by the source modules. Based on input source parameters such as emittances, beta functions, energy, energy spread, beam length and beam offsets, the source modules generate particle distributions with defined shapes such as trapezoidal, Gaussian or flattop.

In the multi-stage acceleration example illustrated by Fig.~\ref{fig:ABEL_structure}, there is one source that generate the main beam to be tracked through the accelerator chain. The driver source generates a drive beam for every plasma stage, which may then be tracked through a radio-frequency (RF) linac before being passed to the plasma stages.

More realistic source simulation modules, such as using ASTRA, may be added in the future.

\subsection{Plasma acceleration stage}
After the drive beam and main beam is passed to a plasma acceleration stage, the beams are evolved using models with varying degrees of fidelity. At the most basic level, the main beam is simply evolved by solving Hill's equation with homogeneous energy gains. 

At intermediate fidelity and speed, ABEL contains a transverse instability stage module which adopts a quasi-static approach, where one single time step of Wake-T simulation is used to calculate the plasma wake, which is assumed unchanged for the rest of the propagation through the stage. The plasma wake is then used as input in the built-in transverse-instability model \cite{Stupakov_short-range_wakefunction, Chen_modeling_2020, my_thesis} and ion motion model \cite{Benedetti_ion_motion_2017, Mehrling_ion_motion_2018, Burov_ionMotion}. In addition, this module also includes radiation reaction \cite{Deng_radiation_reaction_2012}.

The current stage module with the highest fidelity evolves the incoming beams using the PIC code HiPACE++. As HiPACE++ is a highly parallelised PIC-code optimised for GPUs, this module is suited for performing calculations on HPC-clusters. The HiPACE++ module configures all HiPACE++ input files and cluster job scripts, converts the input beams to the format being used by HiPACE++ and submits the simulation job to the HPC-cluster.

All stage modules also support plasma density ramps.

\subsection{Interstage}
All implemented interstage modules applies the self-correction mechanism \cite{Lindstrom_self-correcting_2021} to improve the timing tolerance and optimise the main beam current profile to reduce the energy spread. The basic interstage module applies the self-correction mechanism to the main beam directly through coordinate transformations. 

More advanced interstage modules track the main beam through a compact lattice using codes such as ELEGANT and ImpactX. The lattice consists of dipoles, non-linear plasma lenses, chicane dipoles and a sextupole designed to correct chromaticity locally, cancel dispersion up to third order, preserve emittance and also allows for non-invasive in and out-coupling of drivers \cite{Drobniak_nonlinear_plasma_lens_2025}.

\subsection{Other machine components}
In addition to the outlined beamline elements, ABEL also has modules for RF acceleration structures (basic and CLIC\-opti), combiner rings, damping rings, beam delivery systems and collider IPs (calculations performed using GUINEA-PIG).

\section{SIMULATION CAPABILITIES}
Each beam-tracking simulation in ABEL are performed as shots similar to experiments, allowing the user to include stochastic shot-to-shot imperfections such as beam position, temporal and energy jitter. Simulation runs can be structured as single shot, multi-shot, or as parameter scans and optimisation.

\subsection{Single-shot, multi-shot and scan}
As the name suggests, single-shot simulations only track the beam through the defined beamline once. Multi-shot simulations track beams through multiple copies of the same machine, but allow for introducing stochastic imperfections between each shot. Furthermore, ABEL has the capability of performing automated multi-dimensional parameter scans, as demonstrated in the use case section. Scans also support multiple shots per scan step, which can be performed in parallel.

\subsection{Optimisation and cost model}
ABEL supports multi-dimensional parameter optimisation using Bayesian optimisation to explore a user-defined parameter space and automatically tunes the parameters based on the defined merit function. Combining low-fidelity models and a cost model \cite{Lindstrom_IPAC_2025} that takes into account the full programme cost consisting of the construction cost, integrated power cost\footnote{Required to collect e.g.\ $\SI{1}{\atto\barn^{-1}}$ of data, hence taking machine performance such as luminosity into account.}, maintenance cost and carbon tax, ABEL was recently used to optimise the baseline update for HALHF \cite{Foster_HALHFv2}.

\section{USE CASE: DRIVE BEAM TRANSVERSE JITTER TOLERANCE FOR HALHF}
\begin{figure}[tb]
    \centering
    \begin{subfigure}[b]{\columnwidth}
        \centering
        \includegraphics[width=0.99\columnwidth]{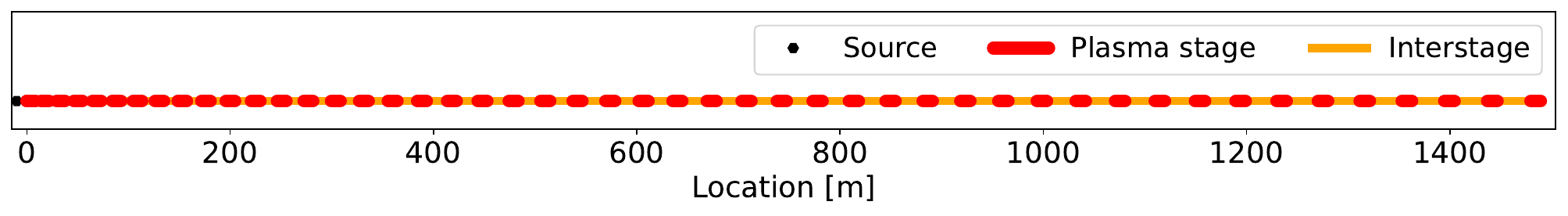}
        \label{fig:sub1}
    \end{subfigure}


    \begin{subfigure}[b]{\columnwidth}
        \centering
        \includegraphics[width=0.99\columnwidth]{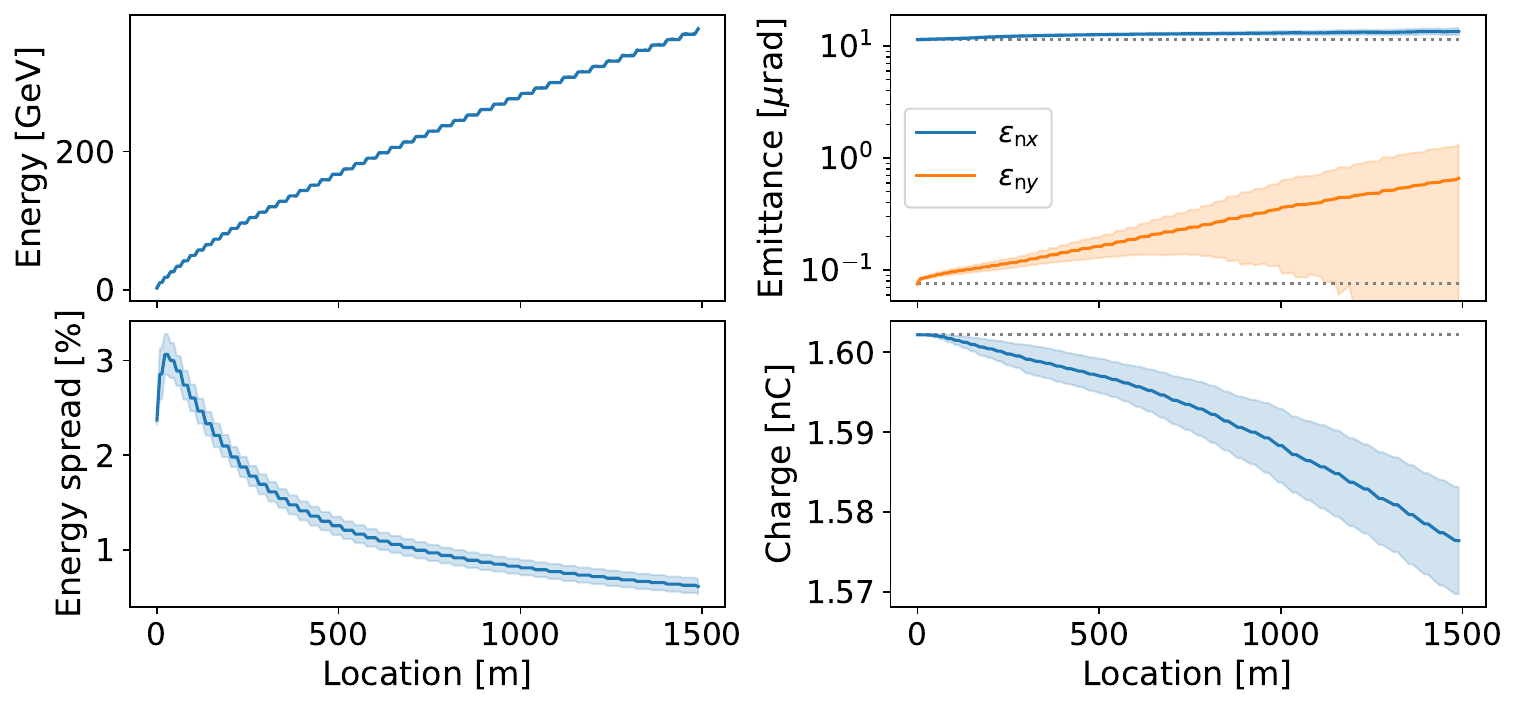}
    \end{subfigure}

    \caption{Main beam parameter evolution along the plasma linac at \SI{100}{\nano\meter} drive beam jitter. Ion motion enabled. An outline of the linac is also included in the top panel.\vspace{-0.3cm}}
    \label{fig:HALHFv2_100nm-jitter_evolution}
\end{figure}

To illustrate the capabilities of ABEL, we present selected results of drive beam jitter tolerance studies for the PWFA linac of the recently proposed HALHF particle collider \cite{Foster_HALHFv1, Foster_HALHFv2}. This study employs ABEL's built-in beam dynamics model that incorporates transverse instability, ion motion (assuming helium ions) and radiation reaction to track and evolve an electron beam through all the 48 plasma acceleration stages and interstages of the plasma linac outlined in Ref.~\cite{Foster_HALHFv2}. An outline of the linac elements is shown in the top panel of Fig.~\ref{fig:HALHFv2_100nm-jitter_evolution}. At each plasma stage, the drive beam is injected with a random transverse offset in both the $x$- and $y$-direction that is determined by the transverse jitter. Hence, such jitter will repeatedly seed the transverse instability at each plasma stage and may accumulate into severe emittance growth and charge losses if unmitigated. Here we will introduce a moderate amount of ion motion and examine its damping effect at various levels of beam jitter, and make a comparison with results with no ion motion.

Furthermore, each plasma stage also has plasma density up- and downramps for matching the main beam beta functions. The self-correction mechanism is applied through the interstages, which are modelled using the ELEGANT interstage module.

The tolerance study is set up as a scan over multiple levels of drive beam jitter, with 10 shots per jitter level. The evolution of beam energy, relative energy spread, normalised emittances $\varepsilon_{\mathrm{n}x,y}$ and beam charge of 10 shots performed at \SI{100}{\nano\meter} drive beam jitter with ion motion are shown in Fig.~\ref{fig:HALHFv2_100nm-jitter_evolution}. The solid lines mark the mean values of the 10 shots, while the error is calculated from the standard deviation. The electron beam in this case is accelerated from \SI{3}{\giga\electronvolt} to \SI{376}{\giga\electronvolt}. Self-correction suppressed the energy spread to sub-percent levels, and only a small fraction of charge was lost. $\varepsilon_{\mathrm{n}x}$ only increased slightly, while $\varepsilon_{\mathrm{n}y}$ may grow to the \SI{}{\milli\meter\milli\radian} level.

\begin{figure}[tb]
    \centering
    \includegraphics[width=0.99\columnwidth]{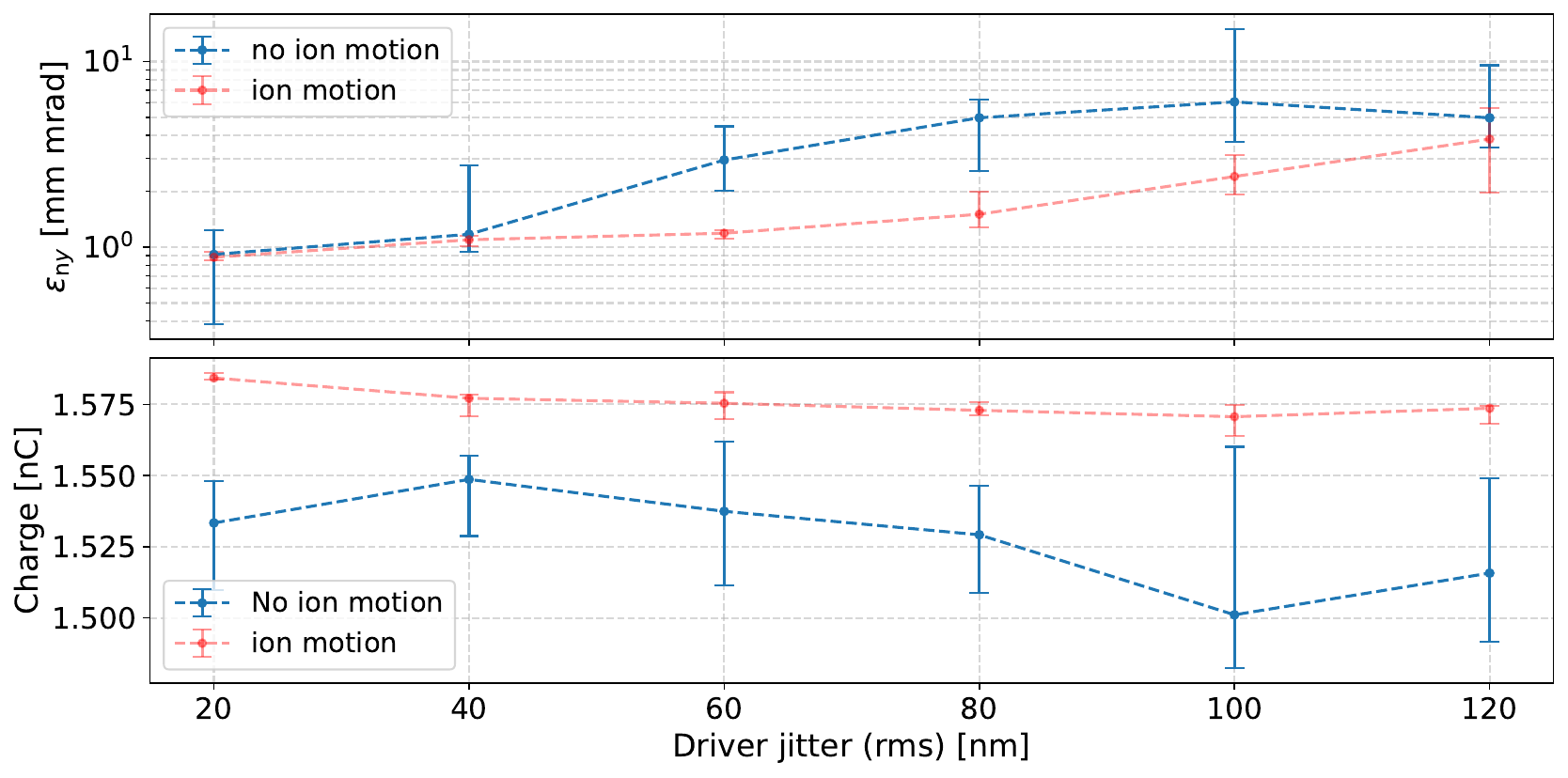}
    \caption{Vertical normalised emittance and beam charge of the output main beam. Multiple levels of drive beam jitter were introduced in the scan, and beam tracking were repeated 10 times at each jitter level. The median value of each step are plotted together with the first and third quartiles. A scan without ion motion is also included. \vspace{-0.5cm}}
    \label{fig:HALHFv2_emitt_scan}
\end{figure}

The output values of main beam charge and $\varepsilon_{\mathrm{n}y}$ are measured for all shots and grouped together based on jitter level to calculate the median values for each group. Error bars are added to mark the first and third quartiles of the group datasets in Fig.~\ref{fig:HALHFv2_emitt_scan}. To illustrate the damping effect of ion motion, we have also included the results of a scan excluding ion motion from the shots. In summary, ion motion reduced median $\varepsilon_{\mathrm{n}y}$ by factors of 1.04--3.31 and increased resulting beam charge by a few percent. The results for the output main beam normalised horizontal emittance are not included, but median $\varepsilon_{\mathrm{n}x}$ for the ion motion cases were consistently lower across all jitter levels by factors $\sim 2$. Further uses of the ABEL framework to study transverse tolerances can be found in Ref.~\cite{Adli_IPAC_2025}.

\vspace{-0.2cm}
\section{CONCLUSION}
The modular approach of ABEL seamlessly integrates models and specialised codes through openPMD and thus allows for flexible simulation runs of entire machines with desired accuracy and speed. ABEL supports automated multi-dimensional parameter scans and possesses an extensive suite of diagnostics, as well as possessing automated multi-dimensional parameter optimisation capabilities linked to a global cost model. Due to its modular structure, ABEL is readily adaptable to other applications, including FELs, strong-field QED experiments, and accelerator test facilities. Hence, ABEL is positioned as a well-suited tool for consistent global project optimisation.

\vspace{-0.05cm}
\section{ACKNOWLEDGEMENTS}
This work was supported by the Research Council of Norway (NFR Grant No. 313770) and the European Union (ERC, 101116161). The computations were performed on resources provided by Sigma2 - the National Infrastructure for High Performance Computing and 
Data Storage in Norway.


%
%
\ifboolexpr{bool{jacowbiblatex}}%
	{\printbibliography}%
	{%

} 
%
%


\end{document}